# A statistical-thermodynamic analysis of stably ordered substitutional structures in graphene

Taras M. Radchenko[*], Valentyn A. Tatarenko

*Dept. of Solid State Theory, G. V. Kurdyumov Institute for Metal Physics, N.A.S.U.,*

*36 Academician Vernadsky Blvd., UA-03680 Kyyiv-142, Ukraine*

**Abstract**

Ordered distributions of carbon and substitutional dopant ($A$) atoms over the sites of a graphene lattice, i.e. $C_mA$ superstructures with dopant contents $c = 1/(m+1)$, and problem of their stability are considered theoretically. The ranges of values of interatomic-interaction parameters providing the low-temperature stability of the graphene-based $C_7A$, $C_3A$, and $CA$ superstructures are determined within the framework of both the third-nearest-neighbor Ising model and, more realistically, the all-coordination-shell interaction model. The first model results in the "omission" (instability) of some predicted superstructures, while the second model shows that all predicted superstructures are stable at the certain values of interatomic-interaction energies. Even short-range interatomic interactions provide a stability of some superstructures, while only long-range interactions stabilize others.

**1. Introduction**

Graphene is a hot-topic object in both materials science and condensed matter physics, where it is a popular model system for investigations [1, 2]. In a crystal lattice of graphene (so-called two-dimensional carbon), atoms are distributed over the vertices of regular hexagons, as shown in Fig. 1. Due to a high mechanical strength, hardness, heat conductivity [3], and electrical conductivity, graphene is a prospective material for the wide applications in the different fields: from nanoelectronics, where graphene is a strong candidate for replacing silicon in the integrated circuit chips, to coating of airliner fuselages [4, 5].

Dopant atoms in graphene can improve some of its physical properties for a wider range of applications. Particularly, dopant atoms change the band structure strongly dependent on atomic order and, consequently, provide a tool to control electrical conductivity of this material [2, 6–19].

The statistical-thermodynamic and kinetic models of a long-range atomic order (LRO) [20] for a two-dimensional graphene-based crystal lattice have been constructed in [21, 22]. However, a topical problem of thermodynamic stability of graphene-based structures has not been raised and the ranges of values of interatomic-interaction energies (parameters) providing their stability were not determined. Therefore, this work is focused on the results of applying of the low-temperature stability conditions [23–26] for the graphene-based (super)structures.

---
[*] Corresponding author. Tel.: +380 44 424 12 21; fax: +380 44 424 25 61.
E-mail address: taras.radchenko@gmail.com (T. M. Radchenko)

## 2. Model

The crystal lattice of graphene (see Fig. 1) consists of two interpenetrating hexagonal sublattices displaced with respect to each other by the vector $\mathbf{h} = \mathbf{a}_1/3 + 2\mathbf{a}_2/3$, where $\mathbf{a}_1$ and $\mathbf{a}_2$ are the basis translation vectors of the lattice along [1 0] and [0 1] directions, respectively, in the oblique system of coordinates. As Fig. 1 shows, each $ABCD$ primitive unit cell contains two sites, and each site position, $\mathbf{r}$, is described by sum of two vectors: $\mathbf{r} = \mathbf{R} + \mathbf{h}_q$, where the vector $\mathbf{R}$ refers to the unit cell's origin position, $q = 1, 2$ is the number of sublattice, $\mathbf{h}_q$ is the distance of the $q$-th type site within the cell with respect to the origin. Since origin of the coordinate system coincides with one of the lattice sites, it is easy to see from Fig. 1 that $\mathbf{h}_1 = \mathbf{0}$ and $\mathbf{h}_2 = \mathbf{h}$. The vector $\mathbf{R}$ relates to the basis vectors as $\mathbf{R} = n_1\mathbf{a}_1 + n_2\mathbf{a}_2$, where $n_1$ and $n_2$ are integers. The radii of the first ten coordination shells of the lattice in Fig. 1 are $r_1 = a_0$, $r_2 = \sqrt{3}a_0$, $r_3 = 2a_0$, $r_4 = \sqrt{7}a_0$, $r_5 = 3a_0$, $r_6 = 2\sqrt{3}a_0$, $r_7 = \sqrt{13}a_0$, $r_8 = 4a_0$, $r_9 = \sqrt{19}a_0$, $r_{10} = \sqrt{21}a_0$, where $a_0$ is a distance between the nearest-neighbor sites.

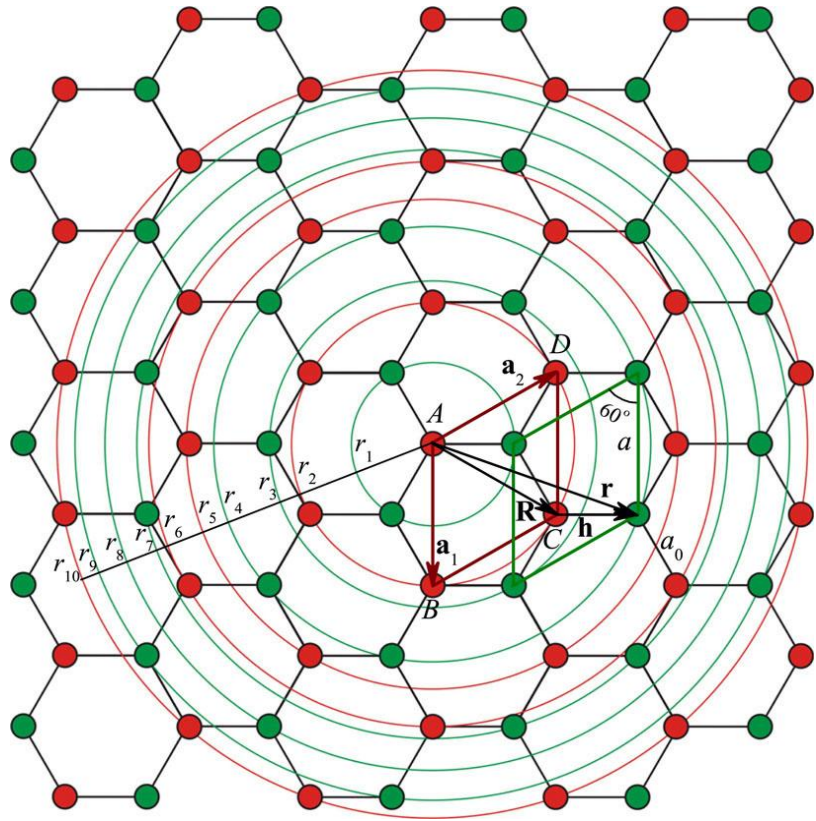

Fig. 1. Crystal lattice of graphene. Here, $ABCD$ is a primitive unit cell; $\mathbf{a}_1$ and $\mathbf{a}_2$ are the basis translation vectors of the lattice; $a$ is the lattice translation parameter; $a_0$ is a distance between the nearest-neighbor sites; circles (with radii $r_1$, $r_2$, ..., $r_{10}$) denote the first ten coordination shells (zones) with respect to the origin (at $A$ site) of the oblique coordinate system.

*2.1. Interatomic-interaction energies*

We will use a model of pairwise interactions [20] between carbon and substitutional dopant (denoted by *A*) atoms in the graphene-type lattice. Let us denote $W_{pq}^{CC}(\mathbf{R} - \mathbf{R}')$, $W_{pq}^{AA}(\mathbf{R} - \mathbf{R}')$, $W_{pq}^{CA}(\mathbf{R} - \mathbf{R}')$ as pairwise interaction energies of C–C, A–A, C–A pairs of atoms, respectively, situated at the sites of *p*-th and *q*-th ($p, q = 1, 2$) sublattices within the unit cells with origins ("zero" sites) at $\mathbf{R}$ and $\mathbf{R}'$. These three variables define so-called mixing energy (in the literature, other names are "ordering energy" and "interchange energy") [20, 23–30],

$$w_{pq}(\mathbf{R} - \mathbf{R}') \equiv W_{pq}^{CC}(\mathbf{R} - \mathbf{R}') + W_{pq}^{AA}(\mathbf{R} - \mathbf{R}') - 2W_{pq}^{CA}(\mathbf{R} - \mathbf{R}'). \tag{1}$$

The mixing energy in the form of Eq. (1) is used often for the analysis of the equilibrium atomic order [20, 23–26] as well as the ordering kinetics [20, 27–30]. It will be shown below (Eq. (8)) that, in the self-consistent field (mean-field) approximation [20], $w_{pq}(\mathbf{R} - \mathbf{R}')$ defines the configurational internal energy of the system.

For the statistical-thermodynamic description of the interatomic interactions in all coordination shells, or arbitrary-range interactions, it is conveniently to apply the Fourier transform [20]

$$\tilde{w}_{pq}(\mathbf{k}) \equiv \sum_{\mathbf{R}} w_{pq}(\mathbf{R} - \mathbf{R}')e^{-i\mathbf{k}\cdot(\mathbf{R}-\mathbf{R}')}, \tag{2}$$

where $\mathbf{k}$ is a wave vector belonging to the reciprocal space of graphene lattice (see Fig. 2) and generating a certain superstructure, and the summation is over all unit cell's origin positions, $\mathbf{R}$. Since $p, q = 1, 2$, the variables $\tilde{w}_{pq}(\mathbf{k})$ form the 2×2 dimension mixing-energy matrix [20]:

$$\begin{pmatrix} \tilde{w}_{11}(\mathbf{k}) & \tilde{w}_{12}(\mathbf{k}) \\ \tilde{w}_{12}^*(\mathbf{k}) & \tilde{w}_{11}(\mathbf{k}) \end{pmatrix}, \tag{3}$$

where $\tilde{w}_{12}^*(\mathbf{k})$ is a complex conjugate to $\tilde{w}_{12}(\mathbf{k})$ (subscripts referred to the numbers of sublattices), and symmetry relations $\tilde{w}_{11}(\mathbf{k}) = \tilde{w}_{22}(\mathbf{k})$, $\tilde{w}_{21}(\mathbf{k}) = \tilde{w}_{12}^*(\mathbf{k})$ are taken into account [20].

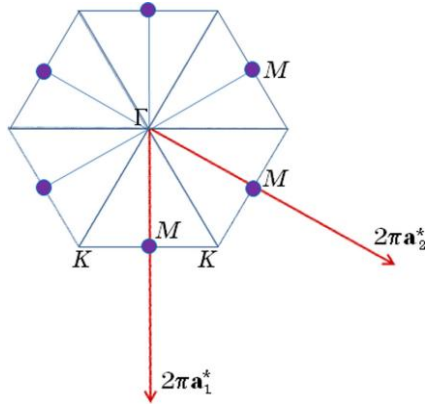

Fig. 2. The first Brillouin zone of the reciprocal space of graphene lattice, where $\Gamma$, $M$, $K$ are its high-symmetry points, and $\mathbf{a}_1^*$, $\mathbf{a}_2^*$ are the basis translation vectors of the two-dimensional reciprocal lattice.

The mixing-energy matrix (3) has two eigenvalues, $\lambda_\sigma(\mathbf{k})$, and eigenvectors, $\mathbf{v}_\sigma(\mathbf{k})$ [20]:

$$\lambda_1(\mathbf{k}) = \tilde{w}_{11}(\mathbf{k}) + |\tilde{w}_{12}(\mathbf{k})|, \qquad \lambda_2(\mathbf{k}) = \tilde{w}_{11}(\mathbf{k}) - |\tilde{w}_{12}(\mathbf{k})|, \qquad (4a)$$

$$\mathbf{v}_1(\mathbf{k}) = \frac{1}{\sqrt{2}}\begin{pmatrix} 1 \\ \tilde{w}_{12}^*(\mathbf{k})/|\tilde{w}_{12}(\mathbf{k})| \end{pmatrix}, \qquad \mathbf{v}_2(\mathbf{k}) = \frac{1}{\sqrt{2}}\begin{pmatrix} 1 \\ -\tilde{w}_{12}^*(\mathbf{k})/|\tilde{w}_{12}(\mathbf{k})| \end{pmatrix}. \qquad (4b)$$

Any reciprocal-space vector $\mathbf{k}$ has an expansion [20]

$$\mathbf{k} = (k_x; k_y) = 2\pi(k_1 \mathbf{a}_1^* + k_2 \mathbf{a}_2^*) = \{k_1, k_2\}, \qquad (5)$$

where $\mathbf{a}_1^*$ and $\mathbf{a}_2^*$ are the reciprocal-lattice translation vectors along the [1 0] and [0 1] directions, respectively, related to the lattice translation parameter as $|\mathbf{a}_1^*| = |\mathbf{a}_2^*| = 1/a$.

Using Eq. (2) and assuming the central-symmetric (isotropic) interatomic interactions, one can write expressions for the elements of the mixing-energy matrix (3):

$$\tilde{w}_{11}(\mathbf{k}) \cong w_2\left[ e^{i2\pi k_1} + e^{-i2\pi k_1} + e^{i2\pi k_2} + e^{-i2\pi k_2} + e^{i2\pi(k_1+k_2)} + e^{-i2\pi(k_1+k_2)} \right] +$$

$$+ w_5\left[ e^{i2\pi(k_1-k_2)} + e^{-i2\pi(k_1-k_2)} + e^{i2\pi(2k_1+k_2)} + e^{-i2\pi(2k_1+k_2)} + e^{i2\pi(k_1+2k_2)} + e^{-i2\pi(k_1+2k_2)} \right] +$$

$$+ w_6\left[ e^{i4\pi k_1} + e^{-i4\pi k_1} + e^{i4\pi k_2} + e^{-i4\pi k_2} + e^{i4\pi(k_1+k_2)} + e^{-i4\pi(k_1+k_2)} \right] +$$

$$+ w_{10}\left[ e^{i2\pi(k_1-2k_2)} + e^{-i2\pi(k_1-2k_2)} + e^{i2\pi(k_1+3k_2)} + e^{-i2\pi(k_1+3k_2)} + e^{i2\pi(2k_1-k_2)} + e^{-i2\pi(2k_1-k_2)} +\right.$$

$$\left. + e^{i2\pi(2k_1+3k_2)} + e^{-i2\pi(2k_1+3k_2)} + e^{i2\pi(3k_1+k_2)} + e^{-i2\pi(3k_1+k_2)} + e^{i2\pi(3k_1+2k_2)} + e^{-i2\pi(3k_1+2k_2)} \right] + \ldots, \qquad (6a)$$

$$\tilde{w}_{12}(\mathbf{k}) \cong w_1\left[ 1 + e^{i2\pi k_2} + e^{i2\pi(k_1+k_2)} \right] + w_3\left[ e^{i2\pi k_1} + e^{-i2\pi k_1} + e^{i2\pi(k_1+2k_2)} \right] +$$

Table 1. Expressions for the elements of the mixing-energy matrix (3) and its eigenvalues for the super-structure ("ordering") wave vectors of Γ, M, and K stars (see Fig. 2).

| Star | **k** | $\tilde{w}_{11}(\mathbf{k})$ | $\tilde{w}_{12}(\mathbf{k})$ | $\lambda_1(\mathbf{k})$ | $\lambda_2(\mathbf{k})$ |
|---|---|---|---|---|---|
| Γ | **0** | $6w_2 + 6w_5 + 6w_6 + 12w_{10} + ...$ | $3w_1 + 3w_3 + 6w_4 + 6w_7 + 3w_8 + 6w_9 + ...$ | $6w_2 + 6w_5 + 6w_6 + 12w_{10} + $ $+|3w_1 + 3w_3 + 6w_4 + $ $+ 6w_7 + 3w_8 + 6w_9| + ...$ | $6w_2 + 6w_5 + 6w_6 + 12w_{10} - $ $-|3w_1 + 3w_3 + 6w_4 + $ $+ 6w_7 + 3w_8 + 6w_9| + ...$ |
| M | $\mathbf{k}_1^M = \left\{\dfrac{1}{2}, 0\right\}$ | $-2w_2 - 2w_5 + 6w_6 - 4w_{10} + ...$ | $w_1 - 3w_3 + 2w_4 + 2w_7 - 3w_8 + 2w_9 + ...$ | $-2w_2 - 2w_5 + 6w_6 - 4w_{10} + $ $+|w_1 - 3w_3 + 2w_4 + $ $+ 2w_7 - 3w_8 + 2w_9| + ...$ | $-2w_2 - 2w_5 + 6w_6 - 4w_{10} - $ $-|w_1 - 3w_3 + 2w_4 + $ $+ 2w_7 - 3w_8 + 2w_9| + ...$ |
| M | $\mathbf{k}_2^M = \left\{0, \dfrac{1}{2}\right\}$ | $-2w_2 - 2w_5 + 6w_6 - 4w_{10} + ...$ | $-w_1 + 3w_3 - 2w_4 - 2w_7 + 3w_8 - 2w_9 + ...$ | $-2w_2 - 2w_5 + 6w_6 - 4w_{10} + $ $+|-w_1 + 3w_3 - 2w_4 - $ $- 2w_7 + 3w_8 - 2w_9| + ...$ | $-2w_2 - 2w_5 + 6w_6 - 4w_{10} - $ $-|-w_1 + 3w_3 - 2w_4 - $ $- 2w_7 + 3w_8 - 2w_9| + ...$ |
| M | $\mathbf{k}_3^M = \left\{-\dfrac{1}{2}, \dfrac{1}{2}\right\}$ | $-2w_2 - 2w_5 + 6w_6 - 4w_{10} + ...$ | $w_1 - 3w_3 + 2w_4 + 2w_7 - 3w_8 + 2w_9 + ...$ | $-2w_2 - 2w_5 + 6w_6 - 4w_{10} + $ $+|w_1 - 3w_3 + 2w_4 + $ $+ 2w_7 - 3w_8 + 2w_9| + ...$ | $-2w_2 - 2w_5 + 6w_6 - 4w_{10} - $ $-|w_1 - 3w_3 + 2w_4 + $ $+ 2w_7 - 3w_8 + 2w_9| + ...$ |
| K | $\mathbf{k}_1^K = \left\{\dfrac{2}{3}, -\dfrac{1}{3}\right\}$ | $-3w_2 + 6w_5 - 3w_6 - 6w_{10} + ...$ | 0 | $-3w_2 + 6w_5 - 3w_6 - 6w_{10} + ...$ | $-3w_2 + 6w_5 - 3w_6 - 6w_{10} + ...$ |
| K | $\mathbf{k}_2^K = \left\{\dfrac{1}{3}, \dfrac{1}{3}\right\}$ | $-3w_2 + 6w_5 - 3w_6 - 6w_{10} + ...$ | 0 | $-3w_2 + 6w_5 - 3w_6 - 6w_{10} + ...$ | $-3w_2 + 6w_5 - 3w_6 - 6w_{10} + ...$ |

$$+ w_4 \left[ e^{-i2\pi(k_1+k_2)} + e^{-i2\pi k_2} + e^{i2\pi(2k_1+k_2)} + e^{i4\pi(k_1+k_2)} + e^{i4\pi k_2} + e^{-i2\pi(k_1-k_2)} \right] +$$

$$+ w_7 \left[ e^{i4\pi k_1} + e^{-i4\pi k_1} + e^{-i2\pi(2k_1+k_2)} + e^{i2\pi(k_1-k_2)} + e^{i2\pi(2k_1+3k_2)} + e^{i2\pi(k_1+3k_2)} \right] +$$

$$+ w_8 \left[ e^{-i2\pi(k_1+2k_2)} + e^{i2\pi(3k_1+2k_2)} + e^{-i2\pi(k_1-2k_2)} \right] +$$

$$+ w_9 \left[ e^{-i4\pi(k_1+k_2)} + e^{-i4\pi k_2} + e^{i2\pi(3k_1+k_2)} + e^{i6\pi(k_1+k_2)} + e^{i6\pi k_2} + e^{-i2\pi(2k_1-k_2)} \right] + ..., \quad (6b)$$

where $w_1, w_2, ..., w_{10}$ are the first-, second-, ..., tenth-neighbor mixing energies, respectively.

Eqs. (2) and (4a) yield expressions for the elements of the mixing-energy matrix (3) and its eigenvalues in the high-symmetry points of the first Brillouin zone (see Fig. 2 and Table 1).

## 2.2. Configurational free energies

The total free energy of a structure is [20, 23–29]

$$F = U - TS, \quad (7)$$

where, in the self-consistent field (mean-field) approximation [20, 23–29],

$$U \cong \frac{1}{2} \sum_{p,q=1}^{2} \sum_{\mathbf{R},\mathbf{R}'} w_{pq}(\mathbf{R} - \mathbf{R}') P_p(\mathbf{R}) P_q(\mathbf{R}') \quad (8)$$

is a configuration-dependent part of the internal energy and

$$S = -k_B \sum_{q=1}^{2} \sum_{\mathbf{R}'} \left[ P_q(\mathbf{R}') \ln P_q(\mathbf{R}') + (1 - P_q(\mathbf{R}')) \ln(1 - P_q(\mathbf{R}')) \right] \quad (9)$$

is a configuration-dependent part of the entropy. Here, $T$ is a temperature; $w_{pq}(\mathbf{R} - \mathbf{R}')$ is given by Eq. (1); $P_q(\mathbf{R})$ is a probability of finding a dopant ($A$) atom at the site of the $q$-th subblatice with unit-cell's origin position at $\mathbf{R}$ (see Fig. 1); $k_B$ denotes the Boltzmann constant. We neglected the vacancies, assuming that $P_q^C(\mathbf{R}) + P_q^A(\mathbf{R}) \equiv 1$. The summation in Eqs. (8) and (9) is over all unit-cell's origin positions ($\mathbf{R}$, $\mathbf{R}'$) and all sublattices, i.e. over all two-dimensional Ising lattice sites.

The ordered distributions of C and substitutional dopant $A$-atoms over the sites of the graphene lattice at the stoichiometries 1/4 ($C_3A$), 1/2 ($CA$), and 1/8 ($C_7A$) are shown in Figs. 3–5. Using the static concentration waves method [20], the single-site occupation probability functions, $P_q(\mathbf{R})$, for different (non)stoichimetric compositions are obtained in [21, 22]. In a completely ordered state, at $T = 0$ K, $P_q(\mathbf{R}) = 1$ or 0; in a completely disordered state, $P_q(\mathbf{R}) = c$ — an atomic fraction of dopant atoms. Substituting $P_q(\mathbf{R})$ into Eqs. (8), (9), one can write the configurational free energies for different structures.

Configurational free energies (per atom) for $C_3A$ superstructures in Figs. 3a, 3b, 3c are, respectively,

$$F_1 \cong \frac{1}{2} c^2 \lambda_1(\mathbf{0}) + \frac{3}{32} \left( \eta_2^I \right)^2 \lambda_2(\mathbf{k}_1^M) - TS_1(c, \eta_2^I), \quad (10a)$$

$$F_2 \cong \frac{1}{2} c^2 \lambda_1(\mathbf{0}) + \frac{1}{32} \left[ 2 \left( \eta_1^{II} \right)^2 \lambda_1(\mathbf{k}_1^M) + \left( \eta_2^{II} \right)^2 \lambda_2(\mathbf{k}_1^M) \right] - TS_2(c, \eta_1^{II}, \eta_2^{II}), \quad (10b)$$

$$F_3 \cong \frac{1}{2} c^2 \lambda_1(\mathbf{0}) + \frac{1}{32} \left[ \left( \eta_0^{III} \right)^2 \lambda_2(\mathbf{0}) + \left( \eta_1^{III} \right)^2 \lambda_1(\mathbf{k}_1^M) + \left( \eta_2^{III} \right)^2 \lambda_2(\mathbf{k}_1^M) \right] - TS_3(c, \eta_0^{III}, \eta_1^{III}, \eta_2^{III}), \quad (10c)$$

where $\eta_\sigma^\aleph$ ($\sigma = 0$, 1 or 2) are the LRO parameters ($\aleph$ index denotes their total number for a given structure; $\aleph = $ I, II or III), $0 \leq \eta_\sigma^\aleph \leq 1$.

Configurational free energies (per atom) for $CA$ superstructures in Figs. 4a, 4b, 4c are, respectively,

$$F_1 \cong \frac{1}{2} c^2 \lambda_1(\mathbf{0}) + \frac{1}{8} \left( \eta_1^I \right)^2 \lambda_1(\mathbf{k}_1^M) - TS_1(c, \eta_1^I), \quad (11a)$$

$$F_2 \cong \frac{1}{2} c^2 \lambda_1(\mathbf{0}) + \frac{1}{8} \left( \eta_2^I \right)^2 \lambda_2(\mathbf{k}_1^M) - TS_2(c, \eta_2^I), \quad (11b)$$

$$F_3 \cong \frac{1}{2} c^2 \lambda_1(\mathbf{0}) + \frac{1}{8} \left( \eta_0^I \right)^2 \lambda_2(\mathbf{0}) - TS_3(c, \eta_0^I). \quad (11c)$$

Configurational free energy (per atom) for $C_7A$ superstructure (Fig. 5) is

$$F \cong \frac{1}{2}c^2\lambda_1(\mathbf{0}) + \frac{1}{128}\left[\left(\eta_0^{III}\right)^2\lambda_2(\mathbf{0}) + 3\left(\eta_1^{III}\right)^2\lambda_1(\mathbf{k}_1^M) + 3\left(\eta_2^{III}\right)^2\lambda_2(\mathbf{k}_1^M)\right] - TS(c,\eta_0^{III},\eta_1^{III},\eta_2^{III}). \quad (12)$$

Interatomic-interaction parameters, $\lambda_1(\mathbf{0})$, $\lambda_2(\mathbf{0})$, $\lambda_1(\mathbf{k}_1^M)$, $\lambda_2(\mathbf{k}_1^M)$, entering into the configurational free energies (10)–(12) are given in Table 1.

As follows from Eq. (7), the low-temperature (i.e. at $T \approx 0$ K) stability of a structure, when contribu-tion of the entropy, $S$, to the total free energy, $F$, is small, depends on the internal energy, $U$ [20, 23–26]. At $T \approx 0$ K, the stable is a phase which has the lowest internal energy as compared with other phases of the same composition [20, 23–26] (here we are neglecting the possibility of the formation of mechanical mixture of pure components and different structures). So to calculate the low-temperature stability ranges, we used the following stability condition [20, 23–26]: minimum of the configurational free energy $F = U|_{T=0}$, setting in Eqs. (10)–(12) $T = 0$ and $\eta_\sigma^N = 1$. This minimum is realized in a certain range of the interatomic-interaction parameters entering into Eqs. (10)–(12).

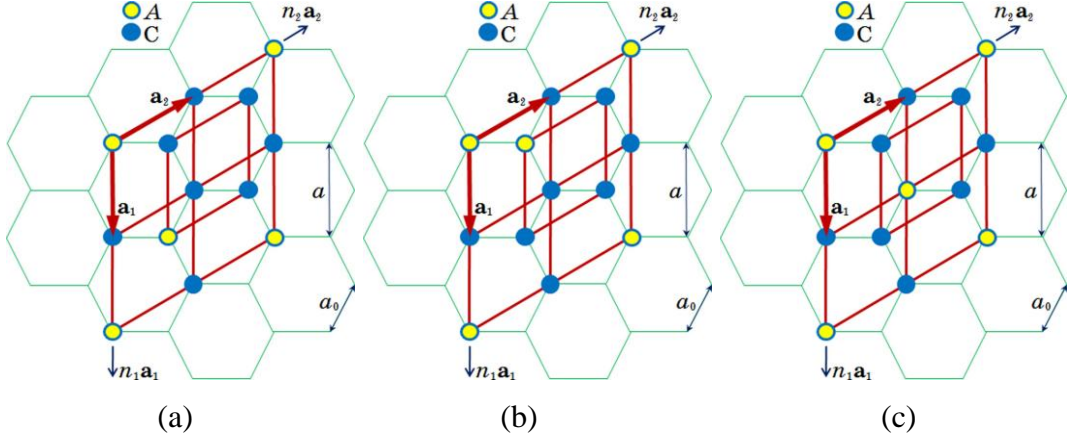

Fig. 3. Primitive unit cells of the graphene-based $C_3A$ superstructures described by one (a), two (b), and three (c) LRO parameters.

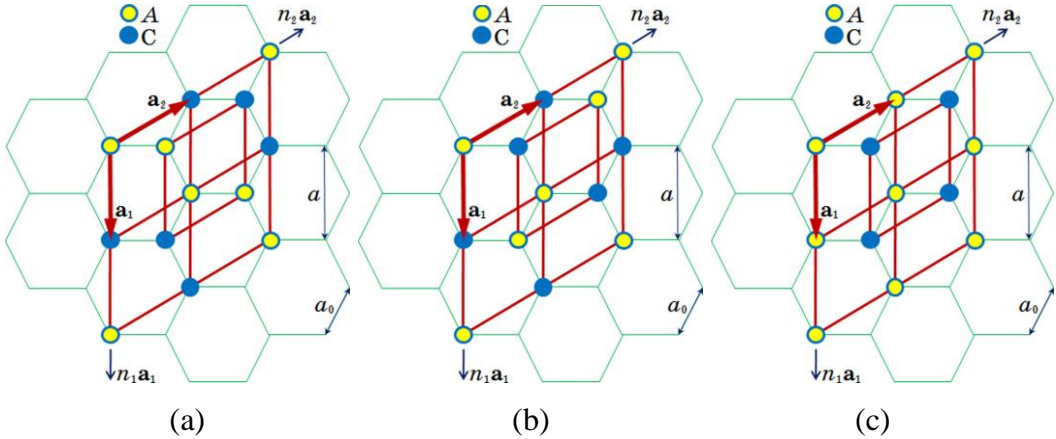

Fig. 4. Different atomic arrangements in the primitive unit cells of the graphene-based $CA$ superstructures described by one LRO parameter.

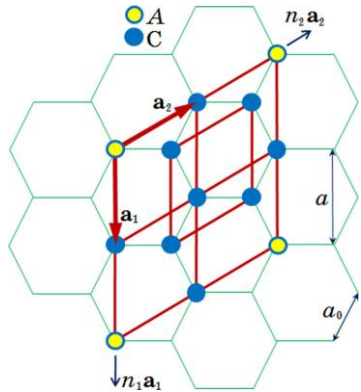

Fig. 5. Primitive unit cell of the graphene-based $C_7A$ superstructure described by three LRO parameters.

## 3. Results

### 3.1. C$_3$A and CA superstructures

These superstructures seem the most interesting, since there are three different ordered distributions of atoms at both 1/4 and 1/2 stoichiometries (Figs. 3, 4). The low-temperature stability ranges for C$_3$A and CA superstructures are presented in Figs. 6, 7. Two cases are considered: firstly, taking into account only first-, second-, and third-neighbor mixing energies ($w_1$, $w_2$, and $w_3$), vanishing mixing energies in other (distant) coordination shells, and, secondly, taking into account mixing energies in all coordination shells.

In Figs. 6, 7, the stability ranges are obtained from the comparison of the configurational free energies ($F_1$, $F_2$, $F_3$) of the phases which have the same composition [20, 23–26]. That is the ranges denoted in Figs. 6, 7 by *1*, *2*, *3* correspond to the following cases: $F_1 < F_2$ and $F_1 < F_3$, $F_2 < F_1$ and $F_2 < F_3$, $F_3 < F_1$ and $F_3 < F_2$, respectively, where $F_1$, $F_2$, $F_3$ are given by Eqs. 10a-10c for C$_3$A-superstructures and Eqs. 11a-11c for CA-superstructures. At the interphase boundaries, *1–2*, *1–3*, *2–3*, corresponding configurational free energies are equal, $F_1 = F_2$, $F_1 = F_3$, $F_2 = F_3$, respectively. There is also a "triple" point, where $F_1 = F_2 = F_3$.

Results of the third-nearest-neighbor model show (see Figs. 6a, 6b) that if the dopant atoms (*A*) are surrounded by only the opposite-type neighbors (C), then C$_3$A superstructures are thermodynamically favorable (stable) at arbitrary values of the mixing energies (excepting particular case, $w_3/w_1 = 1/3$, when $F_1 = F_2 < F_3$, which is caused by the third-nearest-neighbor model). However, C$_3$A superstructure is thermodynamically unfavorable at any values of the mixing energies (excepting the same particular case,

$w_3/w_1 = 1/3$) even if one of the nearest neighbors of dopant atoms is the same-type atom.

Results of the third-nearest-neighbor model for equiatomic composition show (see Figs. 7a, 7b) that some values of the mixing energies provide the low-temperature stability not only for C$A$ superstructure in which dopant atoms are surrounded by only the opposite-type neighbors, but for the superstructure in which dopant atoms are surrounded by one $A$-atom and two C-atoms. However, if the dopant atoms are surround-ed by two the same-type neighbors, the C$A$ superstructure is thermodynamically unfavorable at any values of the mixing energies (excepting the same case, $w_3/w_1 = 1/3$).

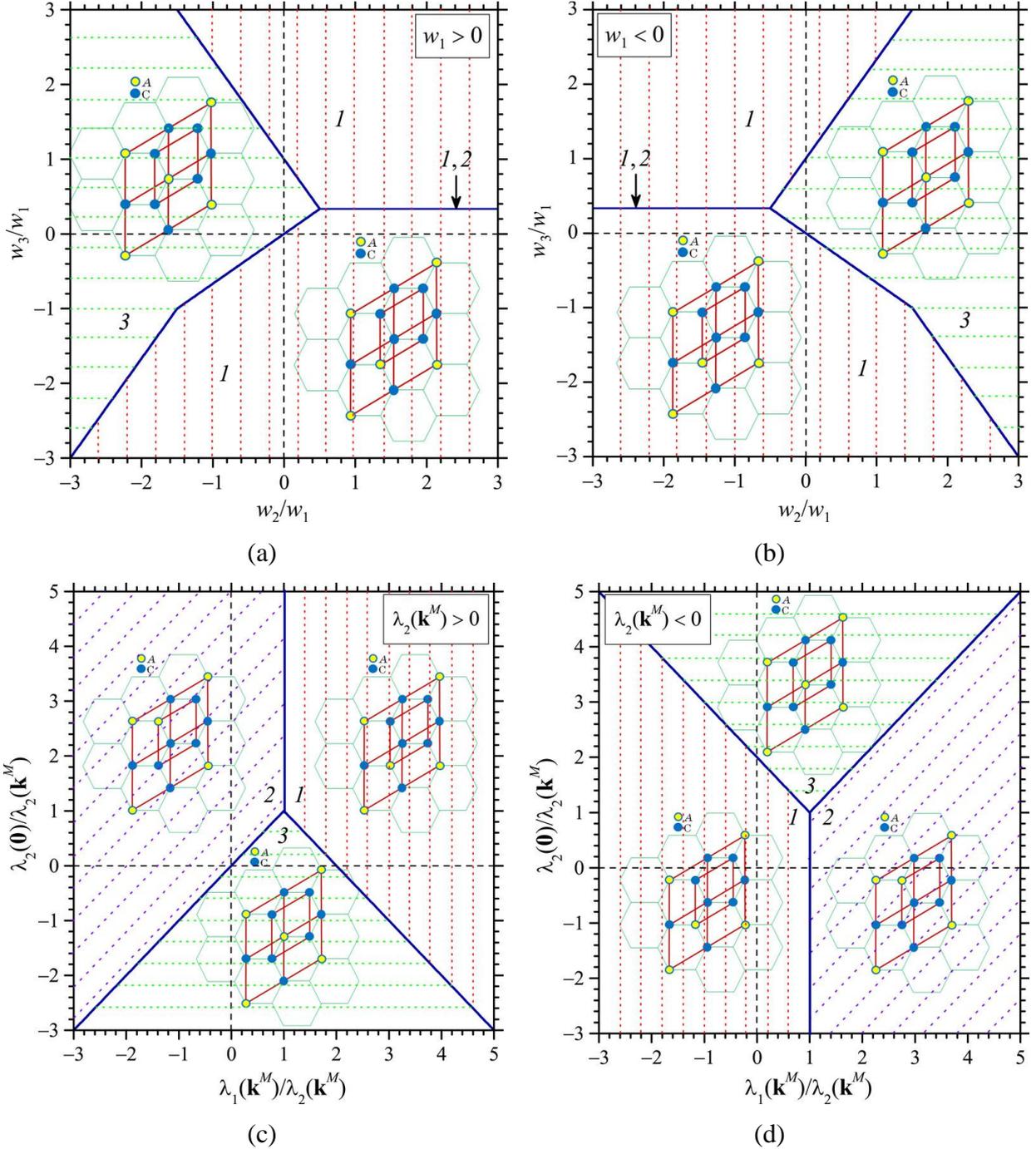

(a) (b)
(c) (d)

Fig. 6. The ranges of values of interatomic-interaction parameters ($w_2/w_1$ and $w_3/w_1$ or $\lambda_1(\mathbf{k}^M)/\lambda_2(\mathbf{k}^M)$ and $\lambda_1(\mathbf{0})/\lambda_2(\mathbf{k}^M)$) providing the low-temperature ($T = 0$ K) stability for C$_3A$ superstructures (see Fig. 3) described by one (*1*), two (*2*) or three (*3*) LRO parameters within the models taking into account interatomic interactions in only three coordination shells (a, b) and in all (c, d) ones.

Thus, an account of the third-nearest-neighbor interatomic interactions always provides the stability for the superstructures (Figs. 3a, 3c, 4c) in which substitutional dopant atoms are surrounded by the opposite-type neighbors. However, an account of (only) these (short-range) interactions can be an inadequate to provide the stability for the superstructures (Fig. 3b, 4a) in which some of the dopant atoms occupy the nearest-neighbor lattice sites.

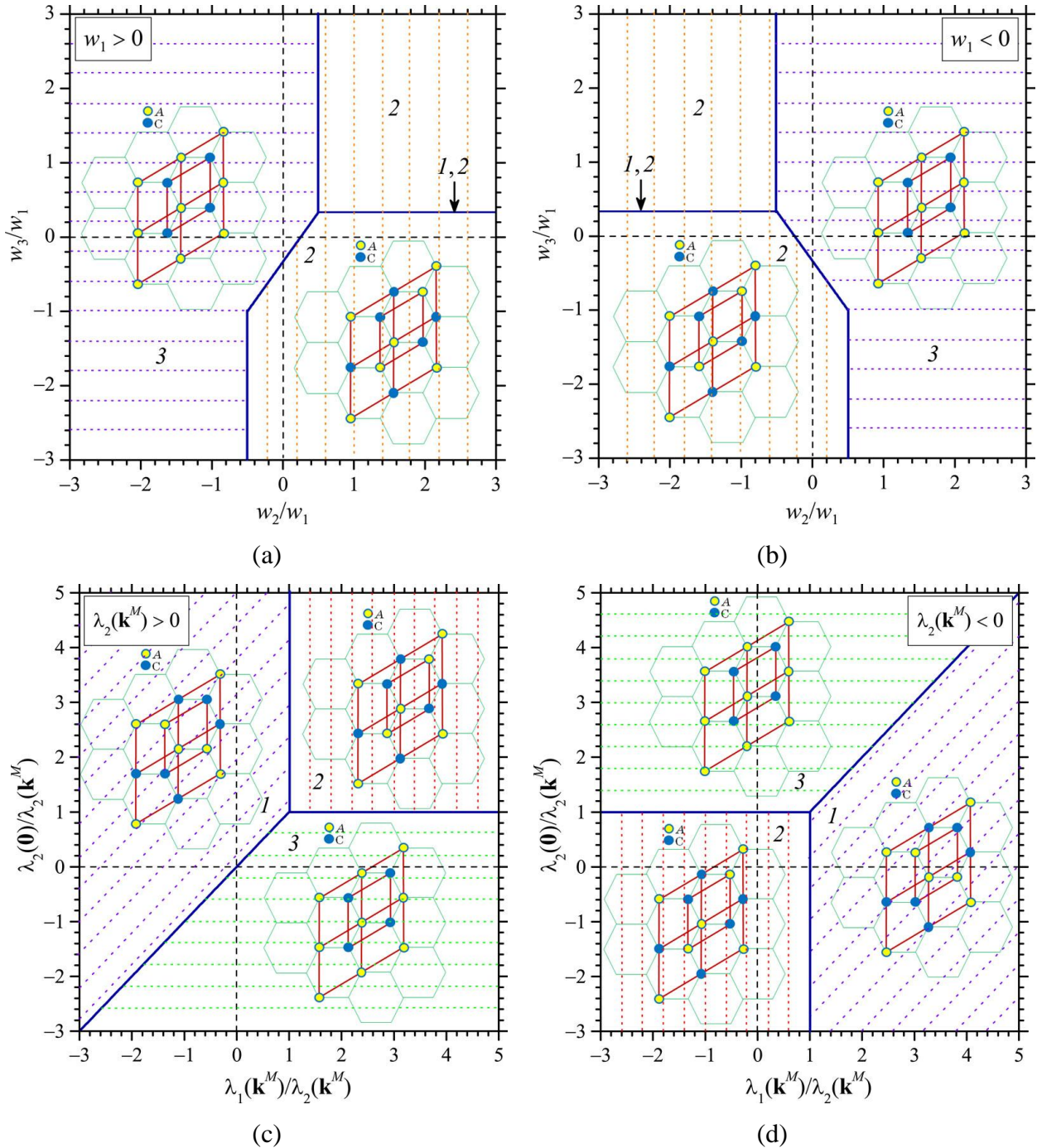

Fig. 7. The same as in the previous figure, but for C$A$ superstructures (see Fig. 4) which are described by one LRO parameter and have different configuration free energy defined with only one interatomic-interactions parameter: $\lambda_1(\mathbf{k}^M)$ (*1*), $\lambda_2(\mathbf{k}^M)$ (*2*) or $\lambda_2(\mathbf{0})$ (*3*).

Figs. 6c, 6d, 7c, 7d demonstrate that accounting the interatomic interactions in all coordination shells yields new results as compared with those obtained within the framework of the third-nearest-neighbor model. Firstly, any of the predicted superstructures (Figs. 4, 5) can be stable at the certain values of interaction energies. Secondly, the reversal of signs for $w_1$ and $\lambda_2(\mathbf{k}^M)$ results in a different symmetry in the disposition of the corresponding stability ranges. In the first case, corresponding stability ranges disposed symmetrically with respect to the ordinate axis (see Figs. 6a and 6b, Figs. 7a and 7b). In the second case, they disposed symmetrically with respect to the "triple" point (1, 1) in which interaction parameters are equal, $\lambda_1(\mathbf{k}^M) = \lambda_2(\mathbf{k}^M) = \lambda_2(\mathbf{0})$ (see Figs. 6c and 6d, Figs. 7c and 7d).

We can compare the ranges of the least values of interatomic-interaction parameters (Fig. 8) with the stability ranges in Figs. 6a, 6b, 7a, 7b, since both ranges are represented in a space of the same parameters, $w_2/w_1$ and $w_3/w_1$. But firstly it should be noted that the $\lambda_2(\mathbf{k}^K)$ doesn't enter into none of the free energies (10)–(12), therefore here we will not consider the range in Fig. 8 where it has a minimum. Comparing Figs. 7a, 7b with Figs. 8a, 8b, one can see that the ranges of minima of $\lambda_2(\mathbf{0})$ and $\lambda_2(\mathbf{k}^M)$ coincide with the stability ranges for C$A$ superstructures, the comparing free energy of which depends on only one of these parameters (the $\lambda_1(\mathbf{0})$ enters in all Eqs. (10)–(12) and "defines" a disordered state of the system).

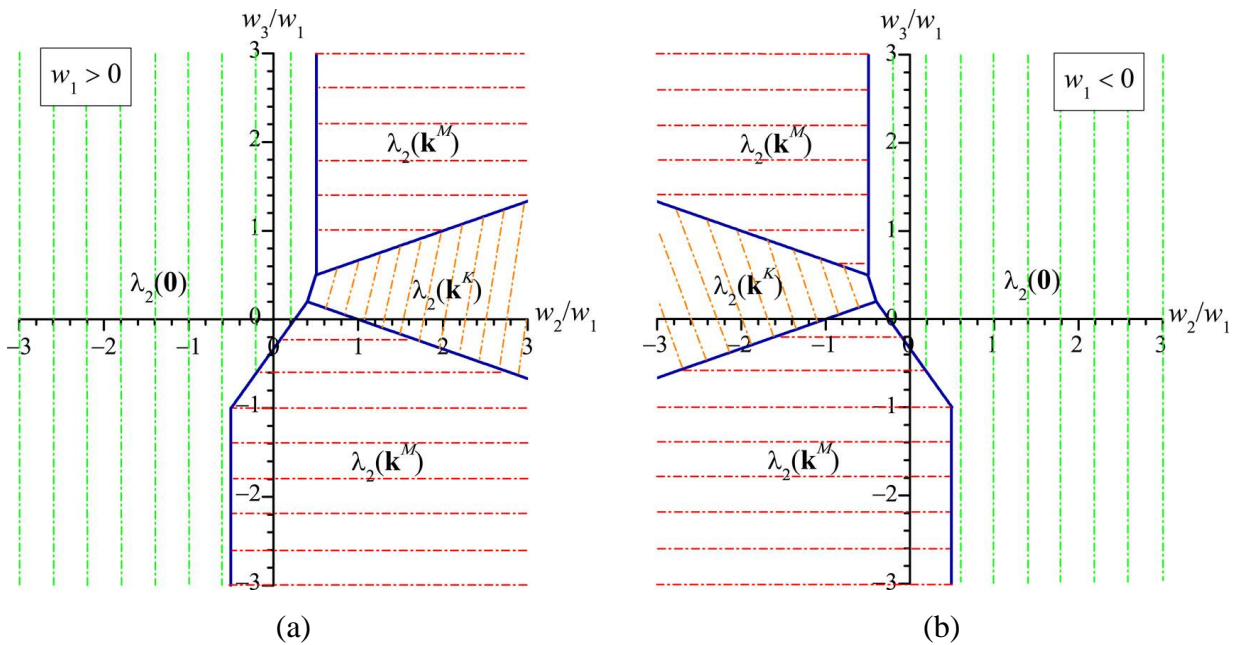

Fig. 8. The ranges of values of interatomic-interaction parameters, $w_2/w_1$ and $w_3/w_1$, providing the least value for one of the ordering thermodynamics parameters: $\lambda_2(\mathbf{0})$, $\lambda_2(\mathbf{k}^M)$ or $\lambda_2(\mathbf{k}^K)$.

Thus, in a given case, when we have to determine a thermodynamic favorable superstructure between those of the same composition and described by only one LRO parameter, a minimum of a certain interatomic-interaction parameter can serve as both necessary and sufficient condition of the stability of corresponding superstructure. However, a minimum of interaction parameter cannot serve neither necessary nor sufficient condition for determining the ranges of the stability (thermodynamic favorableness) for superstructures have the same composition but described by different quantity of the LRO parameters.

*3.2. $C_7A$ superstructure*

At the stoichiometry 1/8, there is only one possible ordered arrangement of carbon and dopant atoms (see Fig. 5). Therefore, at low temperatures, $C_7A$-type superstructure is stable in all space of the values of interatomic-interactions energies.

*3.3. $C_5A$ and $C_2A$ superstructures*

The superstructures with stoichiometric concentrations $c_{st} = 1/6$ ($C_5A$) and $c_{st} = 1/3$ ($C_2A$) are unstable (unrealizable) within the nearest-neighbor interatomic-interactions model because in this case the stability loss temperature [20], $T_0 = -k_B^{-1}c(1-c)\min\lambda_\sigma(\mathbf{k})$, is zero (see rows for $K$-star in Table 1). To investigate the stability problem for the graphene-based superstructures with stoichiometries 1/6 and 1/3, firstly, we have to take into account a long-range (at least, second-neighbor) interatomic interaction, and, secondly, consider more complicated primitive unit cells, for example, those represented in Fig. 9.

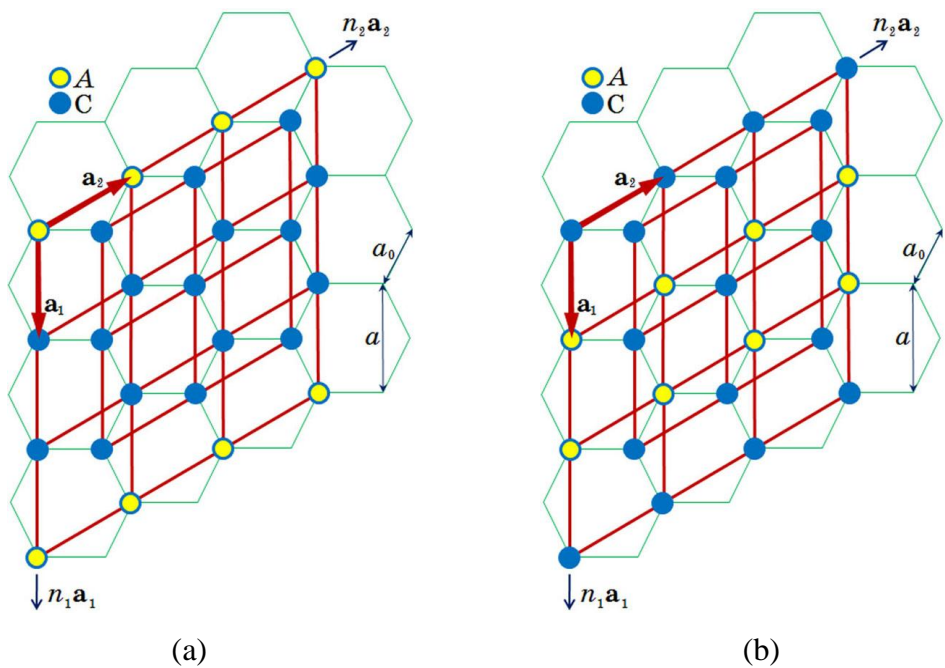

(a)    (b)

Fig. 9. Primitive unit cells of the graphene-based $C_5A$ (a) and $C_2A$ (b) superstructures.

## 4. Conclusions

Using the statistical-thermodynamic approach and applying the phase stability condition, — configuration free energy minimum, — the ranges of values of interatomic-interaction parameters providing the low-temperature stability for the graphene-based $C_3A$, $CA$, and $C_7A$ substitutional superstructures are determined within the scope of both the third-nearest-neighbor interaction model and, more realistically, the all-coordination-shell interaction model. At low temperatures, the latter superstructure can be stable in all space of the values of interatomic-interaction parameters.

On the one hand, even short-range (third-nearest-neighbor) interatomic interactions provide a stability of some superstructures, while on the other hand, within the framework of the third-nearest-neighbor interaction model, the above-mentioned stability condition "reduces" a quantity of possible stable graphene-based superstructures, "forbidding" one of $C_3A$-superstructures and one of $CA$-superstructures in all space of the values of interatomic-interaction parameters. Thereby the third-nearest-neighbor Ising model results in the "omission" (thermodynamic unfavorableness) of some predicted superstructures.

In contrast to the third-nearest-neighbor approach, the account of all number of coordination shells in the interatomic interactions does not result in the "omission" of any of the predicted superstructures, showing thereby that all predicted graphene-based substitutional superstructures are stable at the certain values of interatomic-interaction energies. Moreover, some superstructures ($C_3A$ in Fig. 3b and $CA$ in Fig. 4a) may be stable due to only the long-range interatomic interactions.

Thus, the short-range interatomic interactions always provide stability for the favored in the case of ordering graphene-based superstructures in which the opposite-type neighbors surround all substitutional dopant atoms. However, only long-range interactions may explain the reason for the formation and stability of superstructures in which some dopant atoms occupy the nearest-neighbor lattice sites. Such effect of the long-range interatomic interactions is not surprising, since in three-dimensional structures (alloys, solid solutions), some features (particularly, formation of some superstructures and their stability) can be also explained by the long-range interactions only [20, 23–26]. Moreover, in low-dimensional structures, particularly, in the graphene-based ones, effect of the long-range interatomic interactions may become more "apparent" ("considerable"), as between adatoms in graphene [31], where dramatic enhancement of an electron-mediated interaction, making it long ranged, has been found.

Determination of thermodynamic stable graphene-based superstructure between those of the same composition and described by one LRO parameter reduces to determination of the minimum of a certain interatomic-interaction parameter — necessary and sufficient condition for such superstructure to be stable.

The problem of stability for graphene-based structures is considered at low temperatures. At finite (or room) temperatures, when LRO parameters in Eqs. (8)–(10) are not equal to unity, $\eta_\sigma^\aleph \neq 1$, an entropy contribution to the free energy appears. It will result in a shift of the boundaries between the stability ranges in Figs. 6 and 7, but it will not change the qualitative results, particularly, the long-range interatomic-interactions effect on the stability of the graphene-based (super)structures.

**Acknowledgement**

This work was supported by Grant no. 6Г/30-09 of the National Academy of Sciences of Ukraine for young scientists.